\documentclass[aps,prb,reprint,showpacs,superscriptaddress,footinbib,citeautoscript]{revtex4-1}

\usepackage{graphicx}
\usepackage{amsmath}
\usepackage{amssymb}
\usepackage{bm}
\usepackage{array}
\newcolumntype {s}[1]{@{\hspace{#1}}} 
\usepackage{color}

\usepackage{wrapfig}

\graphicspath{{.}{./EPS/}}

\newcommand* {\ee}{\mathrm{e}}

\newcommand*{\kk}{{\bm{\mathrm{k}}}}
\newcommand*{\rr}{{\bm{\mathrm{r}}}}
\newcommand*{\qq}{{\bm{\mathrm{q}}}}
\newcommand*{\pp}{{\bm{\mathrm{p}}}}

\DeclareMathSymbol{\myRe}{\mathord}{symbols}{"3C}
\renewcommand{\Re}{\myRe\mathrm{e}\,}
\DeclareMathSymbol{\myIm}{\mathord}{symbols}{"3D}

\begin{document}

\title{Suppression of Coulomb exchange energy in quasi-two-dimensional hole systems}

\author{T. Kernreiter}
\affiliation{School of Chemical and Physical Sciences and MacDiarmid
Institute for Advanced Materials and Nanotechnology, Victoria
University of Wellington, PO Box 600, Wellington 6140, New Zealand}

\author{M. Governale}
\affiliation{School of Chemical and Physical Sciences and MacDiarmid
Institute for Advanced Materials and Nanotechnology, Victoria
University of Wellington, PO Box 600, Wellington 6140, New Zealand}

\author{R. Winkler}
\affiliation{Department of Physics, Northern Illinois University,
DeKalb, Illinois 60115, USA}
\affiliation{Materials Science Division, Argonne National
Laboratory, Argonne, Illinois 60439, USA}

\author{U. Z\"ulicke}
\affiliation{School of Chemical and Physical Sciences and MacDiarmid
Institute for Advanced Materials and Nanotechnology, Victoria
University of Wellington, PO Box 600, Wellington 6140, New Zealand}

\date{\today}

\begin{abstract}

We have calculated the exchange-energy contribution to the total energy of
quasi-two-dimensional hole systems realized by a hard-wall quantum-well
confinement of valence-band states in typical semiconductors. The magnitude
of the exchange energy turns out to be suppressed from the value expected
for analogous conduction-band systems whenever the mixing between
heavy-hole and light-hole components is strong. Our results are obtained
using a very general formalism for calculating the exchange energy of
many-particle systems where single-particle states are spinors. We have
applied this formalism to obtain analytical results for spin-3/2 hole
systems in limiting cases.

\end{abstract}

\pacs{73.21.Fg,	   
          71.45.Gm,  
          71.70.Gm,  
          81.05.Ea	    
          } 

\maketitle

\section{Introduction and overview of main results}

In many cases, Coulomb interactions in many-electron systems can
be accounted for by perturbation theory~\cite{giu05}. This is usually
possible at sufficiently high densities where the single-particle
(kinetic-energy or band-dispersion) contribution to the ground-state
energy dominates. In lowest order, interactions give rise to the Hartree
and exchange-interaction terms. The Hartree term embodies the purely
electrostatic Coulomb potential energy of the electrons. In a uniform
system, the Hartree contribution is cancelled by the neutralizing
background of ionic charges in the solid. What remains is the exchange
term, which we focus on in this work.

Since their experimental realization, quantum-confined systems have
become attractive laboratories for the study of interacting
electrons because interaction effects are enhanced in low spatial
dimensions~\cite{giu05}. Examples include the exchange enhancement
of parameters such as spin susceptibility and effective mass in
quasi-two-dimensional (quasi-2D) conduction-band electron
systems~\cite{smi72, tan89, att02}. Curiously, experiments on
similarly confined \emph{valence-band} (hole) states seem to
indicate the absence of exchange-related renormalizations of
electronic parameters~\cite{pin86, win05a, chi11}. Here we reveal a
possible origin of this suppression of interaction effects in
quasi-2D hole systems: the high effective spin associated with
valence-band states. Peculiar Coulombic effects arising from the
spin-3/2 character of holes have previously been noted for bulk
semiconductors. \cite{com72, sch06, sch11, kyr11} More recent
theoretical studies have focused also on quasi-2D hole gases.
\cite{sch94, che01, ker10, ker13, sch13} Our results provide new
insight into the effect of valence-band mixing on physical
properties of confined holes, and the formalism developed here can
also form the basis for further detailed studies of interaction
phenomena in experimentally realized quasi-2D hole systems.

\begin{figure*}
\includegraphics[width=0.31 \textwidth]{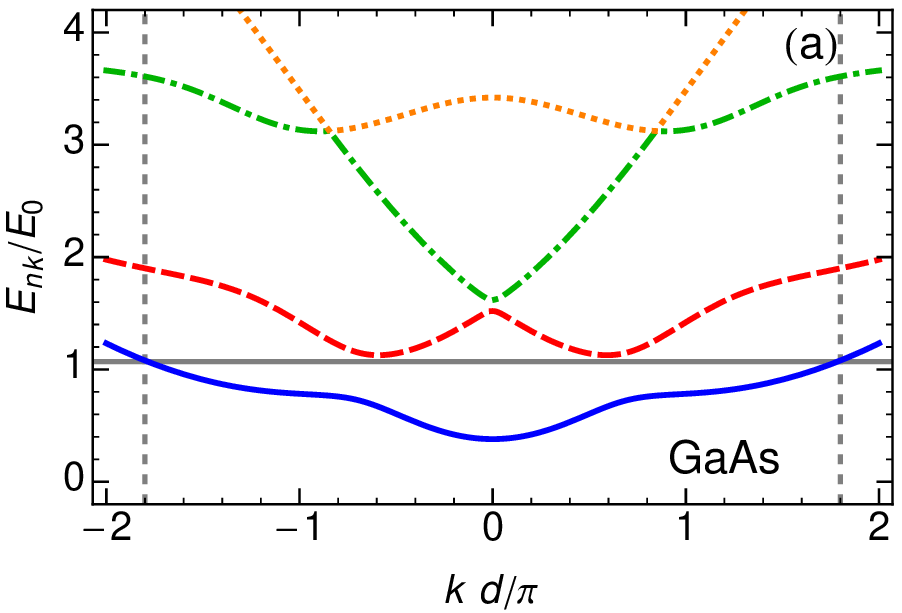}\hfill
\includegraphics[width=0.31 \textwidth]{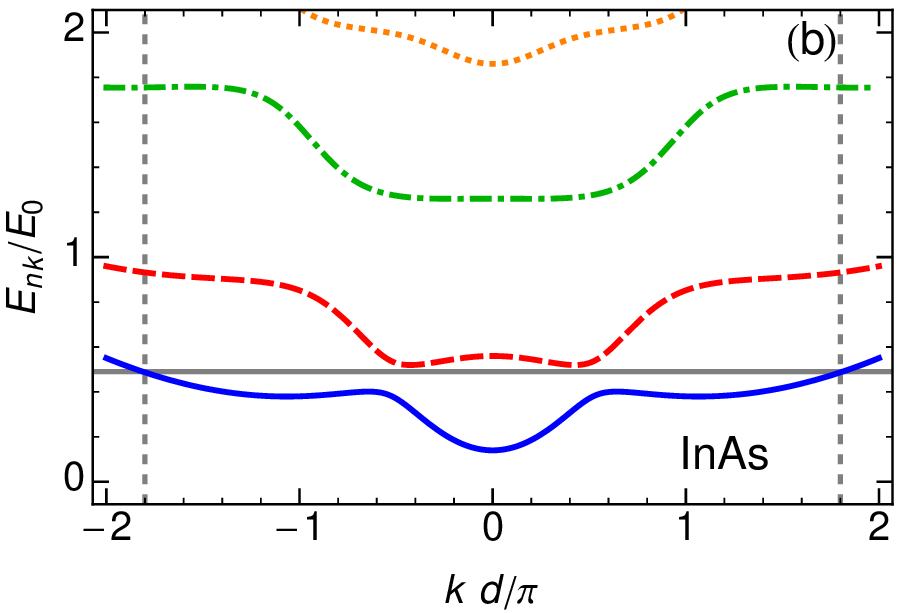}\hfill
\includegraphics[width=0.31 \textwidth]{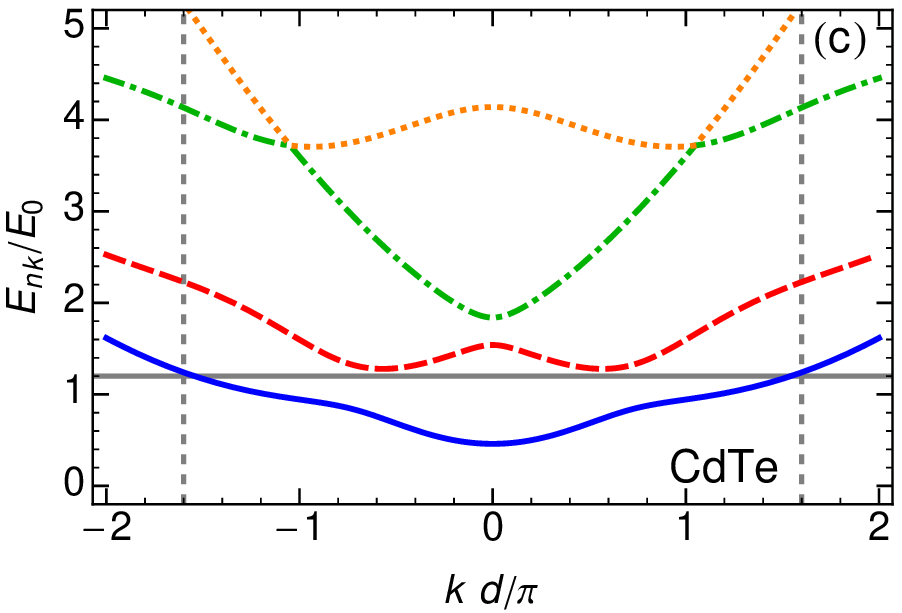}
\caption{\label{fig:Dispersions} Lowest (doubly degenerate) subbands
of 2D hole systems realized by a hard-wall confinement of width $d$,
for (a) GaAs, (b) InAs, and (c) CdTe. $E_0 =\pi^2\hbar^2/ (2m_0d^2)$
is the appropriate size-quantization energy scale. Gray lines
delimit the range of energies and wave vectors for which only the
lowest subband is occupied, which is the regime we focus on in this
work.}
\end{figure*}

The multi-band envelope function approach to electron states in
crystalline solids implies that eigenstates are (in principle,
infinite-dimensional) spinors~\cite{yu10} which affect the matrix
elements describing physical processes in important ways. For an
$n$-fold degenerate band, we may often restrict ourselves to $n$
spinors. The states in the lowest conduction band are usually
non-degenerate, except for spin, so that the spinor structure may
often be ignored. A nontrivial example are hole states in the
topmost valence band of common semiconductors such as Si, Ge, GaAs,
InAs, and CdTe. The \emph{bulk} valence band in these materials is
four-fold degenerate at the band edge, corresponding to an effective
spin 3/2. Away from the band edge, the dispersion splits into the
doubly degenerate heavy-hole (HH) and light-hole (LH) bands. In
quasi-2D systems, the quantum-well confinement results in an energy
splitting between HH and LH bands such that the subbands are
only two-fold degenerate, even at the subband edge with in-plane
wave vector $|\kk|\equiv k=0$. Nevertheless, states in these
subbands need to be described by four-spinors and, except at $k=0$,
these are never pure HHs or LHs. Figure~\ref{fig:Dispersions} shows
the subbands obtained for a hard-wall confinement of holes within
the axial approximation for three semiconductor materials. The
nontrivial physics arises due to the fact that the degeneracy of the
subbands at any fixed value of energy and wave vector $\kk$ is lower
than the number of spinors needed to describe the dynamics of the
Bloch waves. Similar physics applies also to, e.g., 2D electron
systems with Rashba spin-orbit (SO) coupling.~\cite{ras60, byc84,
win03} However, it was found that the exchange energy of such
systems deviates only marginally from that of a simple 2D electron
gas~\cite{che07, che11a}, while the properties of collective excitations
can be more strongly affected by SO coupling~\cite{aga11}.

\begin{figure}[b]
\includegraphics[width=0.8 \columnwidth]{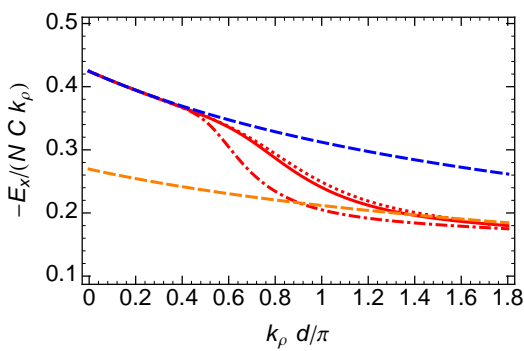}
\caption{\label{fig:Reduction2} Exchange energy per particle
$E_\mathrm{X}/N$ for a hard-wall-confined quasi-2D hole system in
GaAs (solid red curve), InAs (dash-dotted red curve), and CdTe
(dotted red curve). $C \equiv e^2/(4\pi\epsilon\epsilon_0)$ is the
Coulomb-interaction constant, $k_\rho$ denotes the Fermi wave
vector, and $d$ is the quantum-well width. The blue dashed curve
shows the result (\ref{eq:Xfinwidth}) based on the EMA. The orange
dashed curve is $E_\mathrm{X}/N$ for a spin-3/2 hole system
described by a multi-spinor wave function with equal amplitudes for
the heavy-hole and light-hole components.}
\end{figure}

As a reference for our discussion below, we briefly review the
textbook problem~\cite{giu05} of the exchange energy per particle
for a 2D electron gas with zero thickness perpendicular to the
2D plane, giving~\cite{cha71, ste73, giu05}
\begin{equation}\label{eq:Xideal}
\frac{E_\mathrm{X}^{(0)}}{N} = - \frac{4 C}{3\pi}\, k_\rho \equiv 
- \frac{4}{3}\, \sqrt{\frac{2}{\pi}} \; C\,  \sqrt{\rho} =:
\varepsilon_\mathrm{X} \quad .
\end{equation}
Here $N$ is the number of electrons, $C \equiv e^2/(4\pi\epsilon
\epsilon_0)$ is the Coulomb-interaction constant, $\rho$ is the
electron sheet density, and $k_\rho=\sqrt{2\pi\rho}$ denotes the
Fermi wave vector. This result is based on the following
assumptions. (i) The in-plane orbital motion is fully characterized
by plane waves. (ii) We have a two-fold spin degeneracy of the
energy eigenstates, implying that all energy eigenstates can
simultaneously be chosen as eigenstates of spin projection on a
fixed axis. The latter manifests itself in the fact that only
interactions between particles with the same spin projections
contribute to the exchange energy. We emphasize these well-known
points because we find below that these assumptions are not
applicable for quasi-2D hole systems.

Using the same assumptions (i) and (ii) above, one can evaluate the
exchange energy for a \emph{quasi}-2D electron gas in the lowest subband
of a quantum well with hard-wall confinement. Here, $E_\mathrm{X}$
depends also on the well width $d$ and can be written as~\cite{bet94, bet96}
\begin{equation}\label{eq:Xfinwidth}
  \frac{E_\mathrm{X}^{(\mathrm{EMA})}}{N}
  = \varepsilon_\mathrm{X}\,
  \Lambda^{\mathrm{(\mathrm{EMA})}} ( k_\rho d ) \quad .
\end{equation}
The universal function $\Lambda^\mathrm{(\mathrm{EMA})}
(k_\rho d)$ was expressed in Ref.~\onlinecite{bet96} in terms
of a Taylor expansion in its argument. In the following, we will
refer to Eq.\ (\ref{eq:Xfinwidth}) as the effective-mass
approximation (EMA) to the exchange energy. We note that the 
exchange energy in quasi-2D systems is generally reduced with
increasing quantum-well width $d$.

Frequently the exchange energy is expressed in terms of the
dimensionless density parameter $r_s$, which in 2D is the ratio
between the Coulomb energy $C\sqrt{\pi\rho}$ and the kinetic energy
$\pi\rho \, \hbar^2/m^\ast$, assuming a \emph{parabolic} dispersion
$E_\kk = \hbar^2k^2/(2m^\ast)$ with effective mass $m^\ast$. In the
current work, we avoid using the parameter $r_s = m^\ast C /
(\hbar^2\sqrt{\pi\rho})$, the reason being that the systems we study
here do not have a simple parabolic energy dispersion.

\begin{figure}[b]
\includegraphics[width= 0.8 \columnwidth]{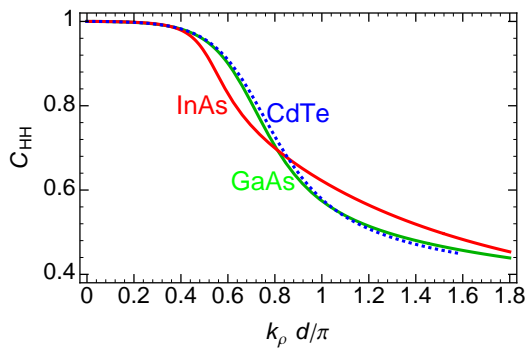}
\caption{\label{fig:HHcont}
Heavy-hole (HH) character of states at the Fermi energy in the
lowest subband, as measured by $C_\mathrm{HH}$, which is the
combined integrated probability density of the HH entries in the
confined-hole spinor wave function. (See text for details.)}
\end{figure}

Using the numerically calculated multi-spinor envelope functions for
quasi-2D hole systems, we evaluate their exchange energy and find it
to deviate from the EMA expression (\ref{eq:Xfinwidth}). As
Fig.~\ref{fig:Reduction2} shows, the EMA behavior is exhibited in
the low-density, small-width limit where the system's states are
essentially HH-like~\cite{win03}. However, at larger densities, even
with still only the lowest quasi-2D subband occupied, the character
of the wave functions becomes mixed, with eventually the HH and LH
components having almost equal weight (see Fig.~\ref{fig:HHcont} and
Ref.~\onlinecite{win03}). The exchange energy turns out to be
suppressed compared to the EMA value as soon as the contributions
from LH components become significant. This can be seen from
Fig.~\ref{fig:HHcont} in conjunction with Fig.~\ref{fig:Reduction2}.
We obtain good agreement between the asymptotic behavior of
$E_\mathrm{X}$ at large densities and an analytical expression
obtained within a simplified model with equally distributed HH and
LH amplitudes. This supports our hypothesis that the suppression of
exchange effects in quasi-2D hole systems arises as a consequence of
valence-band mixing.

In the remainder of this paper, we provide details of the
calculations and further information to illustrate our conclusions. The
basic formalism for calculating the exchange energy of quasi-2D
systems is described in Sec.~\ref{sec:formalism}. In
Sec.~\ref{sec:2Delectrons}, we review known results for quasi-2D
spin-1/2 electron systems that provide a benchmark for comparison
with analogous hole systems. The formalism for calculating the
exchange energy of spin-3/2 quasi-2D hole systems and a discussion
of the obtained results are given in Sec.~\ref{sec:2Dholes}. The
conclusions of our work are summarized in the final Sec.~\ref{sec:conc}.

\section{Exchange energy of quasi-2D systems}
\label{sec:formalism}

\subsection{General formalism}

In its most general form, the many-body Hamiltonian of interacting
band electrons reads
\begin{eqnarray}
  \mathcal{H}& = &\sum_{n, \kk} E_{n\kk}~\hat{c}_{n\kk}^\dagger \hat{c}_{n\kk}
  + \frac{1}{2 A} \sum_{n_i, \kk, \kk', \qq\ne 0} \hspace{-0.5em}
  V^{n_1 n_2}_{n_3 n_4}(\kk, \kk', \qq)
  \nonumber \\ && \hspace{2cm} {} \times
  \hat{c}_{n_1\kk+\qq}^\dagger \hat{c}_{n_2\kk'-\qq}^\dagger
  \hat{c}_{n_3\kk'} \hat{c}_{n_4\kk}\quad ,
\end{eqnarray}
where $A$ is the system volume (i.e., area in the 2D case), and the
operators $\hat{c}_{n\kk}$ ($\hat{c}_{n\kk}^\dagger$) annihilate (create)
an electron with wave vector $\kk$. Note that in the presence of
SO interaction the spin quantum number is, by itself, not a
good quantum number. Thus $n$ is a common index for the orbital
motion in a (sub)band and the spin degree of freedom. \cite{win03}
In the following, we focus on the case of a quantum-well-confined
system, hence the indices $n_i$ are labeling quasi-2D subbands and
$\kk \equiv (k_x, k_y)$ is a 2D vector. We use the jellium model and
therefore the $\mathbf{q} =\mathbf{0}$ component of the interaction
is not present, as it cancels with the external potential due to the
homogeneous positive-charge background that ensures charge
neutrality. Treating the interaction part of the Hamiltonian in
first-order perturbation theory gives the following correction to
the expectation value of the energy of the system:
\begin{equation}
  \label{eq:ex-1}
  \begin{array}[b]{@{}>{\displaystyle}r@{}}
    E_\mathrm{X} = -\frac{1}{2 A}
    \sum_{n_i, \kk, \kk', \qq\ne 0} \hspace{-0.6em}
    V^{n_1 n_2}_{n_3 n_4}(\kk, \kk', \qq)
    \; \langle \hat{c}_{n_1\kk+\qq}^\dagger \hat{c}_{n_3\kk'}\rangle_0
    \\ {} \times
    \langle \hat{c}_{n_2\kk'- \qq}^\dagger \hat{c}_{n_4\kk}\rangle_0\, ,
  \end{array}
\end{equation}
where $\langle \dots\rangle_0$ indicates the expectation value with
respect to the equilibrium density matrix of the non-interacting
system. The correction~(\ref{eq:ex-1}) is the exchange contribution
to the energy. We now notice that
\begin{equation}
  \langle \hat{c}_{n_1 \kk_1}^\dagger \hat{c}_{n_2 \kk_2}\rangle_0
  = \delta_{n_1, n_2}\delta_{\kk_1, \kk_2}\, n_{F}(E_{n_1 \kk_1}),
\end{equation}
where $n_F(E)$ is the Fermi function. Transforming the sums over
wave vectors into integrals and using the relation above, the
exchange energy reads
\begin{eqnarray}
  \label{eq:ex}
  E_\mathrm{X} &=& -\frac{A}{2} \sum_{n, n'}
  \int \! \frac{d^2k}{(2\pi)^2} \int \! \frac{d^2 k'}{(2\pi)^2}
  \; V^{(nn')}_{\kk \kk'}
  \nonumber \\ && \hspace{3cm} \times\,
  n_{F}(E_{n' \kk'}) \; n_{F}(E_{n \kk}) \quad ,
\end{eqnarray}
where we have introduced the abbreviation $V^{(nn')}_{\kk \kk'}
\equiv V^{n' n}_{n' n}(\kk, \kk', \kk'-\kk)$. As a side remark we
can define the Fock self-energy as
\begin{equation}
  \label{eq:selfenergy}
  \Sigma_{n} (\kk) = \sum_{n'}
  \int \!\frac{d^2 k'}{(2\pi)^2}
  \; V^{(nn')}_{\kk \kk'} \; n_{F}(E_{n' \kk'}).
\end{equation}

The exchange energy, Eq.~(\ref{eq:ex}), is an an extensive quantity.
It is customary to present the exchange energy per particle as a
related intensive quantity ($\rho \equiv N/A$),
\begin{subequations}
\begin{align}\label{eq:xchangepp}
  \frac{E_\mathrm{X}}{N} = &-\frac{1}{2\rho}
  \sum_{n, n'} \int \!\frac{d^2 k}{(2\pi)^2} \int \!\frac{d^2 k'}{(2\pi)^2}
  \; V^{(nn')}_{\kk \kk'}
  \nonumber \\ &\hspace{3cm} \times\,
   n_{F}(E_{n' \kk'}) \; n_{F}(E_{n \kk}) \,\, , \\
  \equiv & -\frac{1}{2\rho}
  \sum_{n} \int \!\frac{d^2 k}{(2\pi)^2} \; \Sigma_n(\kk) \; n_{F}(E_{n \kk})
  \quad .\label{eq:xchangepp1}
\end{align}
\end{subequations}
Finally we notice that, in the zero-temperature limit considered here,
the Fermi function becomes $n_{F}(E) =\Theta(E_F - E)$, where
$\Theta(E)$ is the Heavyside step function and $E_F$ the Fermi
energy. Calculation of the quantities given in
Eqs.~(\ref{eq:selfenergy})--(\ref{eq:xchangepp1}) requires explicit
knowledge of the interaction matrix elements. We now discuss their
most general form for a quasi-2D system.

The eigenfunctions of the multi-band envelope-function Hamiltonian
describing a quasi-2D system take the generic form~\cite{win03}
\begin{equation}\label{eq:EFAgen}
  \psi_{n\kk}(\rr, z) =
  \frac{\ee^{i\kk\cdot\rr}}{2\pi}\, \sum_{\nu}
  \xi_{n \kk}^{(\nu)} (z)\: u_\nu (\rr,z) \quad ,
\end{equation}
where $\rr = (x, y)$ is the coordinate in the 2D plane, $\kk$ the
in-plane wave vector, and $\xi_{n \kk}^{(\nu)} (z)$ denotes the
$\nu$th spinor component of the envelope function for the $n$th
subband in the basis of bulk band-edge Bloch functions $u_\nu
(\rr,z)$. Using the general expression (\ref{eq:EFAgen}) for the
electron states, the Coulomb matrix elements are given by
\begin{equation}\label{eq:CoulombGen}
V^{n_1n_2}_{n_3n_4}(\kk, \kk', \qq)
= \frac{2\pi C}{q} \sum_{\nu, \nu'}
F^{n_1n_2}_{n_3 n_4}(\nu, \nu', \kk, \kk', \qq) \quad ,
\end{equation}
with form factors
\begin{widetext}
\begin{equation}\label{eq:genFF}
  F^{n_1n_2}_{n_3 n_4}(\nu, \nu', \kk, \kk', \qq)
  = \int dz \int dz'  \;\ee^{-q |z-z'|}
  \: \xi^{(\nu)*}_{n_1\kk+\qq}(z)
  \, \xi^{(\nu')*}_{n_2\kk'-\qq}(z')
  \, \xi^{(\nu')}_{n_3 \kk'}(z')
  \, \xi^{(\nu)}_{n_4 \kk}(z)
\end{equation}
that take into account the finite quantum-well width as well as the
spinor structure of quasi-2D states. More specifically, we obtain
for the exchange matrix elements
\begin{subequations}
\label{eq:xgen_me}
\begin{equation}\label{eq:xGen}
  V^{(nn')}_{\kk \kk'} = \frac{2\pi C}{|\kk - \kk'|}
  \sum_{\nu, \nu'} F^{(nn')}_{\kk \kk', \nu\nu'} \quad ,
\end{equation}
with form factors
\begin{equation}\label{eq:xgenFF}
  F^{(nn')}_{\kk \kk', \nu\nu'}
  = \int dz \int dz'  \;\ee^{-|\kk - \kk'| |z-z'|}
  \: \xi^{(\nu)*}_{n' \kk'}(z)
  \, \xi^{(\nu')*}_{n \kk}(z')
  \, \xi^{(\nu')}_{n' \kk'}(z')
  \, \xi^{(\nu)}_{n \kk}(z)\quad .
\end{equation}
\end{subequations}
\end{widetext}
Equation (\ref{eq:xchangepp}) combined with Eqs.\ (\ref{eq:xGen})
and (\ref{eq:xgenFF}) provides the most general expression for the
exchange energy of a quasi-2D many-particle system in a multiband
envelope-function formulation. We would like to make the following
remarks concerning Eq.\ (\ref{eq:xgen_me}). (i) The $z$ ($z'$)
integration probes the overlap between spinor components for the
same spinor index $\nu$ ($\nu'$); in the end we sum over $\nu$ and
$\nu'$. In contrast to the usual textbook case, terms with $\nu$
different from $\nu'$ generally contribute to the exchange energy.
(ii) The exchange matrix elements for quasi-2D systems are 
reduced as compared with the corresponding quantities for a
strictly-2D system because $\ee^{-|\kk - \kk'| |z-z'|} < 1$. (iii)
We note the symmetries
\begin{equation}
  \label{eq:form-sym}
  F^{(nn') \, \ast}_{\kk \kk', \nu'\nu} =
  F^{(nn')}_{\kk \kk', \nu\nu'} =
  F^{(n'n)}_{\kk' \kk, \nu'\nu} \quad,
\end{equation}
which have been pointed out in a related context before [see Eq.~(13)
in Ref.~\onlinecite{bet96a}]. The first relation in Eq.\
(\ref{eq:form-sym}) implies that the exchange matrix elements
(\ref{eq:xGen}) are real
\begin{equation}
  V^{(nn')}_{\kk \kk'} = \frac{2\pi C}{|\kk - \kk'|}
  \sum_{\nu, \nu'} \Re F^{(nn')}_{\kk \kk', \nu\nu'} \quad .
\end{equation}
The second relation in Eq.\ (\ref{eq:form-sym}) implies the symmetry
$V^{(nn')}_{\kk \kk'} = V^{(n'n)}_{\kk' \kk}$, which can be useful
to aid efficient numerical computation of Eq.~(\ref{eq:xchangepp}).
(iv) Finally, we point out that the generalized exchange matrix
elements (\ref{eq:xgen_me}) can give interesting physics for a
nontrivial orbital dynamics that can be expressed in spinor form (as
is the case for, e.g., graphene), for a nontrivial $\kk$ dependent
spin texture (as is the case for, e.g., systems with Rashba SO
coupling), as well as for the most general case of coupled multiband
dynamics with SO coupling, where spin and orbital motion are not
separable (as is the case for, e.g., quasi-2D hole systems).

\subsection{Axial Approximation}

In the following, we employ the \emph{axial aproximation\/}, where
the spinor components $\xi_{n \kk}^{(\nu)}(z)$ satisfy
\begin{equation}\label{eq:axWF}
\xi_{n \kk}^{(\nu)}(z) =
\ee^{-i m_\nu \theta} \: \tilde{\xi}_{n k}^{(\nu)}(z) \quad .
\end{equation}
Here we have expressed the 2D wave vector $\kk$ in polar
coordinates, $\kk = (k, \theta)$, and $m_\nu$ is the $z$ projection
of total angular momentum associated with the band edge Bloch
function $u_\nu$. The axial approximation reflects the fact that it
is often possible to group the basis spinors $u_\nu \dot{{}={}} (0,
\dots, 0, 1, 0, \dots, 0)$ into degenerate multiplets associated
with a total angular momentum $j_\nu$ and $z$ projections $m_\nu$.
Within the axial approximation, we find
\begin{equation}\label{eq:xchMat}
  V^{(nn')}_{\kk\kk'} = \frac{2\pi C}{|\kk - \kk'|}
  \sum_{\nu, \nu'} \ee^{i (m_\nu - m_{\nu'})
  (\theta' - \theta)}
  \mathcal{A}^{(\nu, \nu')}_{n' k', n k} (\theta' - \theta) \: ,
\end{equation}
with radial form factors
\begin{widetext}
\begin{equation}\label{eq:overLP}
  \mathcal{A}^{(\nu, \nu')}_{n' k', n k}
  \left(\theta' - \theta\right)
  = \int dz \int dz'  \;\ee^{-|\kk - \kk'| |z-z'|}
   \: \tilde{\xi}^{(\nu)*}_{n' k'}(z)
   \: \tilde{\xi}^{(\nu')*}_{n k}(z')
   \: \tilde{\xi}^{(\nu')}_{n' k'}(z')
   \: \tilde{\xi}^{(\nu)}_{n k}(z) \: . \quad
\end{equation}
Separating out the most important effect of the angular
dependencies, we rewrite Eq.~(\ref{eq:xchMat}) as follows:
\begin{equation} \label{eq:exchange}
  V^{(nn')}_{\kk\kk'}  =  \frac{2\pi C}{|\kk' - \kk|}
  \left\{ \sum_{\nu, \nu'}
    \mathcal{A}^{(\nu, \nu')}_{n' k', n k}
    \left( \theta' - \theta \right)
    - \sum_{\nu, \nu'}
    \left[ 1 - \ee^{i (m_\nu - m_{\nu'})
      (\theta' - \theta)}\right]
    \mathcal{A}^{(\nu, \nu')}_{n' k', n k}
    \left(\theta' - \theta\right)\right\} \quad .
\end{equation}
From the expression (\ref{eq:exchange}) for the Coulomb-interaction
exchange matrix elements, the effect of a nontrivial spinor
structure of the quasi-2D wave function becomes apparent. The first
term on the r.h.s.\ of Eq.~(\ref{eq:exchange}) represents the usual
form-factor renormalization of the exchange energy in a 2D
system~\cite{and82}. The second term only contributes for spinor
wave functions with more than a single nonvanishing component.

Inserting (\ref{eq:exchange}) into the expressions for
the self-energy correction due to exchange eventually yields
\begin{subequations}
\begin{equation}\label{eq:2Dxchange}
  \frac{E_\mathrm{X}}{N} = -\frac{C}{4 \pi^2 \rho}
  \sum_{n n'} \int_0^{k_{\mathrm{F}n}} k \, dk \:
  \int_0^{k_{\mathrm{F}n'}} k' \,  dk'
  \: \int_{0}^{\pi} d\varphi \: \frac{f_{n n'}(k, k', \varphi)}
  {\sqrt{k^2 + k'^2 - 2kk' \cos\varphi}} \quad ,
\end{equation}
where $k_{\mathrm{F}n}$ is the Fermi wave vector of subband $n$,
and the form factors become
\begin{equation}\label{eq:formF}
  f_{n n'}(k, k', \varphi) =
  \sum_{\nu, \nu'}\Re\mathcal{A}^{(\nu, \nu')}_{n' k', n k} (\varphi)
  - 4 \sum_{\nu, \nu' \atop \nu>\nu'}
  \sin^2 \left(\frac{m_\nu - m_{\nu'}}{2} \varphi \right) \:
  \Re \mathcal{A}^{(\nu, \nu')}_{n' k', n k} (\varphi) \quad .
\end{equation}
\end{subequations}
Assuming we have $\aleph$ spin-resolved subbands occupied,
this can be rephrased in the ``dimensionless'' form
\begin{equation}
  \frac{E_\mathrm{X}}{N} = - C \sqrt{\frac{4 \rho}{\pi\aleph^3}}
  \sum_{n=1}^\aleph \, \sum_{n' = 1}^\aleph \,
  \int_0^{\frac{k_{\mathrm{F}n}}{k_\rho}} d\kappa\, \kappa \:
  \int_0^{\frac{k_{\mathrm{F}n'}}{k_\rho}} d\kappa'\, \kappa'
  \: \int_{0}^{\pi} d\varphi \: \frac{f_{n n'}(\kappa k_\rho, \kappa' k_\rho,
  \varphi)}{\sqrt{\kappa^2 + \kappa'^2 - 2\kappa\kappa' \cos\varphi}} \quad ,
\end{equation}
where $k_\rho = \sqrt{\sum_n k_{\mathrm{F}n}^2 / \aleph}\equiv
\sqrt{4\pi\rho/\aleph}$ is a density-related wave vector that
coincides with the Fermi wave vector of a 2D electron system with
$\aleph$-fold (spin-)degenerate eigenenergies.

\subsection{Subband $\kk \cdot \pp$ method}
\label{sec:subkdotp}

To calculate the form factors $\mathcal{A}^{(\nu, \nu')}_{n'
k', n k}$, the explicit form of the spinor functions
$\tilde{\xi}^{(\nu)}_{n k}(z)$ is needed. The subband $\kk \cdot
\pp$ method~\cite{bro85, bro85a} yields these as superpositions
of subband-edge states,
\begin{equation}\label{eq:kdotpExp}
  \tilde{\xi}^{(\nu)}_{n k}(z)
  = \sum_m a^{(\nu, m)}_{n k} \, \tilde{\xi}^{(\nu)}_{m 0}(z)
  \quad ,
\end{equation}
with expansion coefficients $a^{(\nu, m)}_{n k}$. The radial form factors
are then given by
\begin{equation}\label{eq:kdotpFF}
  \mathcal{A}^{(\nu, \nu')}_{n' k', n k}
  \left(\theta' - \theta\right) =
  \sum_{m_i} 
  \bar{\mathcal{A}}^{(\nu, \nu')}_{n' k', n k}
  \: \Phi^{m_1 m_2}_{m_3 m_4}(\nu, \nu', |\kk' - \kk|) \: ,
\end{equation}
with
\begin{equation}
  \bar{\mathcal{A}}^{(\nu, \nu')}_{n' k', n k}
  = a^{(\nu, m_1)\, \ast}_{n' k'} a^{(\nu', m_2)\, \ast}_{n k}
  a^{(\nu', m_3)}_{n' k'} a^{(\nu, m_4)}_{n k}
\end{equation}
and
\begin{equation}
  \label{eq:2ndFF}
    \: \Phi^{m_1 m_2}_{m_3 m_4}(\nu, \nu', q) =
    \int \! dz \int \! dz' \;\ee^{-q |z-z'|} \;
    \tilde{\xi}^{(\nu)*}_{m_1 0}(z) \, 
    \tilde{\xi}^{(\nu')*}_{m_2 0}(z') \, 
    \tilde \xi^{(\nu')}_{m_3 0}(z') \, 
    \tilde \xi^{(\nu)}_{m_4 0}(z) \: .
\end{equation}
For a hard-wall confinement of width $d$, one has
\begin{equation}
  \tilde{\xi}^{(\nu)}_{m 0}(z) 
  = \sqrt{2/d} \, \sin (m \pi \, z /d) \quad ,
\end{equation}
with $m =1, 2, \dots$, independent of the spinor index $\nu$. In
this case, the form factors (\ref{eq:2ndFF}) can be evaluated
explicitly, \cite{bet96} giving (with $s \equiv qd = |\kk - \kk'| \,
d$)
\begin{eqnarray}
  \label{eq:hw_FF}
  \Phi_{m_3 m_4}^{m_1m_2} (s)
  & = & \frac{16 \, m_1 m_2 m_3 m_4 \, \pi^4s^2
  \left\{[(-1)^{m_1+m_4} + (-1)^{m_2+m_3}] \ee^{-s}
    - 1 - (-1)^{m_1+m_2+m_3+m_4}\right\}}
  {[(m_2-m_3)^2 \pi^2+s^2] [(m_2+m_3)^2 \pi^2+s^2]
   [(m_1-m_4)^2 \pi^2+s^2] [(m_1+m_4)^2 \pi^2+s^2]} 
   \nonumber\\[1ex]
   && {} + \frac{s}{(m_2-m_3)^2\pi^2+s^2}
   \left(\delta_{m_1-m_2+m_3-m_4,0} + \delta_{m_1+m_2-m_3-m_4,0}
     - \delta_{m_1-m_2+m_3+m_4,0} - \delta_{m_1+m_2-m_3+m_4,0} \right)
   \nonumber\\[1ex]
   && {} + \frac{s}{(m_2+m_3)^2\pi^2+s^2}
   \left(\delta_{m_1-m_2-m_3+m_4,0} - \delta_{m_1+m_2+m_3-m_4,0}
     - \delta_{m_1-m_2-m_3-m_4,0}\right)~,
 \end{eqnarray}
\end{widetext}
where $\delta_{m,m'}$ denotes the Kronecker symbol. Note that Eq.\
(\ref{eq:hw_FF}) implies that $\Phi_{m_3 m_4}^{m_1m_2}$ vanishes if
$m_1 + m_2 + m_3 + m_4$ is odd.

\section{Results for spin-1/2 2D systems
\label{sec:2Delectrons}}

\subsection{Zero-width 2D systems}

To make connections with previous results, we apply the general
formalism described in the previous Section to a 2D electron system
whose transverse density profile is a delta function in $z$. In this case,
only $\Phi^{11}_{11}(s \rightarrow 0) = 1$ from Eq.~(\ref{eq:hw_FF})
is relevant. The form factor (\ref{eq:formF}) is then only a function of
$\varphi$, and its explicit expression depends on the particular
type of electron system. For example, in a $\aleph$-component system
modeled within the EMA, we can use the component-related quantum
number as the band label. Hence, $\bar{\mathcal{A}}^{(\nu,
\nu')}_{n', n} = \delta_{n \nu} \delta_{n' \nu'} \delta_{\nu \nu'}$
independent of $k$ and $k'$, yielding $f_{n n'}(k, k',\varphi)
\equiv\delta_{n n'}$ and the result
\begin{equation}\label{eq:EMAgen}
\frac{E_{\text{X}}}{N} = - \frac{4}{3}
\sqrt{\frac{4}{\pi\aleph}} \; C \, \rho^{\frac{1}{2}} \quad .
\end{equation}
which specializes to the expression (\ref{eq:Xideal}) for a spin-1/2
system where $\aleph=2$.

In contrast, if the 2D electrons are subject to Rashba SO
coupling~\cite{ras60, byc84, win03}
\begin{equation}
H_\mathrm{R} = \frac{\hbar^2 k_\mathrm{so}}{m}\, \left(\sigma_x
\, k_y - \sigma_y\, k_x\right) \quad ,
\end{equation}
where $k_\mathrm{so}$ measures the strength of the Rashba SO
coupling, the Fermi surface splits into two circles with radii
$k_{\mathrm{F}\pm} = k_\rho\, \sqrt{1 \pm\chi}$, where $\chi
=(\rho_+ - \rho_-)/\rho$ is the density imbalance between the two
Fermi seas. We find
\begin{subequations}
  \begin{eqnarray}
    \bar{\mathcal{A}}^{(\nu, \nu')}_{\pm\pm}
    & = & \bar{\mathcal{A}}^{(\nu, \nu)}_{\pm\mp}
    = 1/4 \quad, \\
    \bar{\mathcal{A}}^{(1/2, -1/2)}_{\pm\mp}
    & = & \bar{\mathcal{A}}^{(-1/2, 1/2)}_{\pm\mp} = -1/4 \quad,
  \end{eqnarray}
  thus
  \begin{eqnarray}
    f_{\pm\pm}(k, k',\varphi) & = & \cos^2 (\varphi/2)  \quad,\\
    f_{\pm\mp}(k, k',\varphi) & = & \sin^2(\varphi/2) \quad.
  \end{eqnarray}
\end{subequations}
The exchange energy is then the sum of intra-band and inter-band
contributions
\begin{equation}
  \label{eq:rash_ex}
  \frac{E_\mathrm{X}^{(\mathrm{R})}}{N} = \varepsilon_\mathrm{X} \left[
    \Lambda_\mathrm{intra}^{(\mathrm{R})}(\chi) +
    \Lambda_\mathrm{inter}^{(\mathrm{R})}(\chi)\right] \quad ,
\end{equation}
with
\begin{subequations}
\begin{equation}
  \Lambda_\mathrm{intra}^{(\mathrm{R})}(\chi) =
    \left(\frac{1+\chi}{2}\right)^{3/2}
    + \left(\frac{1-\chi}{2} \right)^{3/2}
\end{equation}
and
\begin{widetext}
\begin{equation}
  \Lambda_\mathrm{inter}^{(\mathrm{R})}(\chi)  = \frac{3}{8}
  \int_0^{\sqrt{1+\chi}} \!\!\! d\kappa \int_0^{\sqrt{1-\chi}}
  \!\!\! d\kappa' (\kappa+\kappa') \left[ E\left(\frac{4\kappa
      \kappa'}{(\kappa+\kappa')^2}\right) - \frac{(\kappa -
    \kappa')^2}{(\kappa +\kappa')^2}
    K\left(\frac{4\kappa\kappa'}{(\kappa+\kappa')^2}\right) \right]
  \quad .
\end{equation}
\end{widetext}
\end{subequations}
Here $K(\cdot)$ and $E(\cdot)$ denote complete elliptic integrals of
the first and second kind, respectively, as defined in
Ref.~\onlinecite{abr72}. The term in square brackets in Eq.\
(\ref{eq:rash_ex}) expresses the correction caused by Rashba SO
coupling to the simple exchange energy $\varepsilon_\mathrm{X}$ of a
2D electron gas with zero width. Explicit calculation shows that
\begin{equation}\label{eq:ExchRashba}
1 \le \Lambda_\mathrm{intra}^{(\mathrm{R})}(\chi) +
\Lambda_\mathrm{inter}^{(\mathrm{R})}(\chi) < 1.002
\end{equation}
for any value of $\chi$, i.e., the presence of Rashba SO coupling
very slightly enhances (the magnitude of) the exchange energy of the
2D electron gas.~\cite{che07} (See Fig.~3 of
Ref.~\onlinecite{che11a} for a clear illustration.) 
Note that the inequality in Eq.~(\ref{eq:ExchRashba}) holds for
parabolic bands with any values of the effective mass and Rashba
SO-coupling strength. The rather small magnitude of this effect is due
to a subtle interplay between the intra-band and inter-band contributions
for the special case of Rashba SO coupling and does not hold for arbitrary
types of SO coupling.~\cite{che07}

\subsection{Quasi-2D systems: Effect of finite width}
 \label{sec:fullumix}

When a quasi-2D spin-1/2 system without SO coupling is confined by a
hard-wall potential of width $d$, the form factors
$\mathcal{A}^{(\nu, \nu')}_{n' k', n k}(\varphi)$ are given by Eq.\
(\ref{eq:kdotpFF}) with
\begin{equation}\label{eq:phi_rel}
\Phi^{1 1}_{1 1}(s) = \frac{3 s^5+20 \pi^2 s^3+32 \pi^4
\left(e^{-s}-1+s\right)}{(s^3+4\pi^2 s)^2}
\, . \quad
\end{equation}
In the EMA case appropriate for electron systems, using again the
spin projection as band label, we have $f_{n n'}(k, k',\varphi) =
\Phi^{1 1}_{1 1}(s) \,\delta_{n n'}$. This yields the result
(\ref{eq:Xfinwidth}), shown as the blue dashed curve in
Fig.~\ref{fig:Reduction2}. The width-dependent function
$\Lambda^{(\mathrm{EMA})}(x)$ can be expressed as a Taylor series
\begin{subequations}
\label{eq:ema-expand}
\begin{equation}
\Lambda^{(\mathrm{EMA})}(x)
= \sum_{n=0}^\infty \lambda^{(\mathrm{EMA})}_n \, x^n \quad ,
\end{equation}
with coefficients
\begin{widetext}
\begin{equation}
\lambda^{(\mathrm{EMA})}_n
=\frac{3}{4 n!}\,\,\left.\frac{d^{n} \, \Phi^{1 1}_{1 1}(s) }{d s^n}
\right|_{s=0}
\;\int_0^1 d\kappa \,\, \kappa \int_0^1 d\kappa' \: \kappa'
\int_{0}^{\pi} d\varphi
\left(\kappa^2+\kappa'^2-2 \kappa\kappa'\cos\varphi \right)^{(n-1)/2}
\quad .
\end{equation}
\end{widetext}
\end{subequations}
This Taylor expansion is quickly convergent. For an error of less
than $10$\%, going up to $n=4$ is sufficient to describe quasi-2D
systems within the density range $k_\rho d < \sqrt{3} \pi$, where
only the lowest subband is occupied.

\section{Exchange energy of spin-3/2 2D~hole systems}
\label{sec:2Dholes}

\subsection{Theoretical description of 2D hole systems}

We have calculated the exchange contribution to the ground state
energy of quasi-2D hole systems, assuming the common case that the
density is such that only the lowest HH-like band is
occupied.~\footnote{Within our model where a hard-wall confinement
has been assumed, this condition implies $\rho < 10^{12}$~cm${}^{-2}$
for a 15-nm-wide quantum well. Density estimates for real samples will
have to be based on a more accurate, self-consistent modelling of
the confining potential.}
The quantum-well growth direction is assumed to be parallel to the [001]
crystallographic axis, and we considered three different host
materials: GaAs, InAs, and CdTe.

We adopt the 4$\times$4 Luttinger model~\cite{lut56}, including a
potential $V(z)$ that models the confinement in $z$-direction. To be
specific, we assume infinitely high barriers at $z =0$ and $z =d$,
with $d$ the width of the quantum well. Using the eigenstates for
spin projection perpendicular to the 2D plane as basis states such
that $u_{3/2}=(1,0,0,0)$, $u_{1/2}=(0,1,0,0)$, $\dots$, the
Luttinger Hamiltonian for $z\parallel [001]$ is given in matrix
representation by
\begin{equation}
\mathcal{H}_\mathrm{L} =\left(
\begin{array}{cccc}
P+Q& L & M & 0 \\
L^* & P-Q& 0& M \\
M^* & 0 & P-Q& -L \\
0 & M^* & -L^*& P+Q
\end{array}
\right),
\label{eq:LuttHam}
\end{equation}
where
\begin{subequations}
\begin{eqnarray}
P& =&\frac{\hbar^2}{2m_0}\gamma_1 (\kk^2+k_z^2) + V(z)~,\\
Q& =&\frac{\hbar^2}{2m_0}\gamma_2 (\kk^2-2k_z^2)~,\\
L& =&-\frac{\hbar^2}{2m_0}2\sqrt{3}\gamma_3~ k_- \, k_z~,\\
M& =&-\frac{\hbar^2}{2m_0} \frac{\sqrt{3}}{2}
\left[(\gamma_2+\gamma_3) \, k_-^2 + (\gamma_2-\gamma_3) \,k_+^2 \right].
\label{eq:Maxial}
\end{eqnarray}
\end{subequations}
Here $k_\pm = k_x\pm i k_y$ in terms of the components of the
in-plane wave vector $\kk = (k_x,k_y)$, $k_z =-i\partial/\partial
z$, and $\gamma_{1, 2, 3}$ are the ma\-te\-rial-dependent Luttinger
band-structure parameters. Their values for the three semiconductor
materials considered here are given in Table \ref{tab:Lutpara}.

\begin{table}[b]
\caption{\label{tab:Lutpara} Luttinger parameters used in the calculations.}
\tabcolsep 2ex
\begin{tabular}{cccc}
\hline\hline
& $\gamma_1$ & $\gamma_2$ & $\gamma_3$ \\ \hline
GaAs (Ref.~\onlinecite{vur01}) & 6.98  & 2.06  & 2.93\\
InAs (Ref.~\onlinecite{vur01}) &  20. & 8.5 & 9.2 \\
CdTe (Ref.~\onlinecite{die01}) & 4.14  & 1.09 & 1.62 \\  \hline  \hline
\end{tabular}
\end{table}

As the second term in Eq.\ (\ref{eq:Maxial}) [the one proportional
to $(\gamma_2-\gamma_3)$] is generally small, it is often neglected
in the matrix element $M$ of the Hamiltonian (\ref{eq:LuttHam}).
This constitutes the axial approximation~\cite{suz74, tre79, win03}
that we also adopt here. As can be seen from Eq.\ (\ref{eq:LuttHam}),
$\mathcal{H}_\mathrm{L}$ is diagonal for $\kk=0$, which implies that the
corresponding eigenstates are of purely HH or LH character. We
utilize the subband $\kk\cdot\pp$ theory~\cite{bro85, bro85a}, described
in Sec.~\ref{sec:subkdotp} above, to numerically solve the multi-band
Schr\"odinger equation for the confined valence-band states. Using this
method, the subband dispersions $E_{nk}$ and the coefficients
$a^{(\nu,m)}_{n k}$ from Eq.~(\ref{eq:kdotpExp}) are obtained. We
include as many basis functions as are necessary to ensure accuracy
of the lowest few subband dispersions. Fig.~\ref{fig:Dispersions} shows
the quasi-2D subbands obtained for CdTe, GaAs, and InAs.

The eigenspinors corresponding to states in the lowest doubly
degenerate quasi-2D hole subband are used to determine the radial
form factors from Eq.~(\ref{eq:kdotpFF}), with analytical
expressions for the bound-state form factors (\ref{eq:2ndFF}) given
in Eq.~(\ref{eq:hw_FF}). Using the result as input for
Eqs.~(\ref{eq:2Dxchange}) and (\ref{eq:formF}), the exchange energy
of the quasi-2D hole system can be calculated. A modified quadrature
method \cite{bak77, dun77} was applied to treat the singularity in
the integrand of Eq.~(\ref{eq:2Dxchange}).

\subsection{Zero-width limit}

In the zero-width limit, only a single pair of degenerate 2D hole subbands
exist with eigenstates that are of purely HH character, i.e., only the spinor
entries related to the $\pm 3/2$ spin projection quantum numbers are
nonzero. This situation is analogous to the EMA-based description of a
spin-1/2 2D electron system. For example, using the representation where
the two bands are related to eigenstates $(1, 0, 0, 0)$ and $(0, 0, 0, 1)$,
we have no inter-band contribution to the exchange energy, and the
intra-band contribution yields Eq.~(\ref{eq:Xideal}). 

\subsection{Fully HH-LH-mixed limit}

In real quasi-2D hole systems, the states at finite in-plane
momentum are not anymore eigenstates of spin projection
perpendicular to the 2D plane~\cite{win03}. In fact, in the limit of
large kinetic energy of the in-plane motion, the spinors have
approximately equal admixture of HH and LH components. This is
illustrated in Fig.~\ref{fig:HHcont}. In this limit, we can
approximate the true spin-3/2 spinor wave functions from the two
degenerate lowest subbands by assuming them to be of the
form~\footnote{Our \textit{Ansatz\/} is compatible with the basic
symmetries exhibited by hole quantum-well states [L.~C.\ Andreani,
A.~Pasquarello, and F.~Bassani, Phys.\ Rev.\ B~\textbf{36}, 5887
(1987)] and satisfactorily approximates the numerically found
spinor patterns.} $(1, 0, 1, 0) [\sin(\pi z/d)]/\sqrt{d}$ and
$(0, 1, 0, 1) [\sin(\pi z/d)]/\sqrt{d}$. Then we can use the
formalism based on Eqs.~(\ref{eq:2Dxchange}) and (\ref{eq:formF}) to
estimate the exchange energy in such a system. We find for this
fully HH-LH-mixed case
\begin{subequations}
\begin{eqnarray}
f_{\pm\pm}(k, k',\varphi) &=&  \Phi^{1 1}_{1 1}(s)\, \cos^2 \varphi \quad , \\
f_{\pm\mp}(k, k',\varphi) &=&  0 \quad  ,
\label{eq:fpm}
\end{eqnarray}
\end{subequations}
with $\Phi^{11}_{11}$ given in Eq.~(\ref{eq:phi_rel}). The resulting
width dependence of the exchange energy is shown as the orange
dashed curve in Fig.~\ref{fig:Reduction2}. Similar to Eq.\
(\ref{eq:ema-expand}) we can perform the Taylor expansion
\begin{equation}\label{eq:fullMix}
\frac{E_\mathrm{X}^{(\mathrm{mix})}}{N}
= \varepsilon_\mathrm{X}\, \Lambda^{(\mathrm{mix})}
(k_\rho d) \equiv \varepsilon_\mathrm{X}\, \sum_{n=0}^{\infty}
\lambda^{(\mathrm{mix})}_n \left(k_\rho d\right)^n,
\end{equation}
with expansion coefficients
\begin{widetext}
\begin{align}
\lambda^{(\mathrm{mix})}_n
= \frac{3}{8 n!} \left.\frac{d^{n} \, \Phi^{1 1}_{1 1}(s)}{d s^n} \right|_{s=0}
\int_0^1 d\kappa \, \kappa \int_0^1 d\kappa' \, \kappa'
\int_{-\pi}^{\pi} d\varphi \; \cos^2(\varphi)
\left(\kappa^2+\kappa'^2-2 \kappa \kappa'\cos\varphi\right)^{(n-1)/2} \quad .
\end{align}
\end{widetext}
To describe quasi-2D hole systems within the density range, where
only the lowest subband is occupied, and with an error of less than
10\%, it is sufficient to go up to $n=6$.

\subsection{Numerical results for the general case}

The variation of the exchange energy as a function of Fermi wave
vector and quantum-well width is shown in Fig.~\ref{fig:Reduction2}.
In the low-density, small-width limit, the curve coincides with the
plot of the expression (\ref{eq:Xfinwidth}) obtained for the EMA
quasi-2D system. In contrast, the fully mixed case (\ref{eq:fullMix})
provides a very good description in the limit of large densities. The
small deviation that persists in the asymptotic limit arises due to
our assumption of a single sine-wave contribution to the hole bound
state in the full-mixing model considered above.

To further bolster our argument that HH-LH mixing is the origin of
exchange-energy suppression in quasi-2D hole systems, we have
calculated the HH character $C_\mathrm{HH}$ of the states at the
Fermi energy by integrating the combined probability densities for
the $\pm 3/2$ spin projection entries in the corresponding
four-spinor wave function. In terms of the latter's expression as a
superposition of states at $\kk=0$ [see Eq.~(\ref{eq:kdotpExp})], we
find
\begin{equation}
C_\mathrm{HH}\equiv\sum_m\sum_{\nu=\pm 3/2}
\left| a^{(\nu,m)}_{1 k_\rho} \right|^2 \quad .
\end{equation}
The results for GaAs, InAs, and CdTe are shown in
Fig.~\ref{fig:HHcont}. Comparison
with Fig.~\ref{fig:Reduction2} shows that the deviation from the
behavior expected for a simple quasi-2D spin-1/2 electron system
occurs when the hole states at the Fermi energy are no longer of
purely HH character.~\footnote{Even though states deep inside the
Fermi sea are still pure HHs, their pairwise contribution to the
exchange energy gets overwhelmed by that of the much more numerous
fully mixed states.}

\section{Conclusions and Outlook}
\label{sec:conc}

We have calculated the exchange-energy contribution to the total
energy of hard-wall-confined quasi-2D hole systems when only the
lowest subband is occupied. Even though there is only a double
degeneracy of states in this band, the wave functions are
four-spinors associated with the intrinsic spin-3/2 degree of
freedom for valence-band states. At sufficiently low densities
and/or small quantum-well widths when the quasi-2D hole states are
almost purely heavy-hole-like, the behavior resembles that of
ordinary (spin-1/2) quasi-2D electron systems. However, as soon as
the light-hole admixture in the spinors becomes appreciable, the
exchange energy turns out to be suppressed.

Performing calculations for the specific case of a hard-wall
confinement enabled us to obtain analytical expressions for relevant
limiting cases and compare these with the numerically calculated 
results for the exchange energy of quasi-2D hole systems.
However, the general formalism to calculate the exchange energy in
confined multi-band systems that is developed in this work can be
readily adapted to more accurate descriptions~\cite{win03} of hole
quantum wells. Our results provide strong evidence that the
suppression of exchange effects will be a general feature of any
situation where HH-LH mixing is substantial, and we therefore
propose this mechanism as the basic explanation of an experimentally
observed absence of exchange renormalizations in quasi-2D hole
systems~\cite{pin86, win05a, chi11}. Depending on sample details,
this behavior may occur in different parameter regimes, e.g., while
HH-LH mixing is strong for larger densities for a hard-wall
confinement, the fully mixed regime occurs at \emph{low\/} densities
for a density-dependent triangular potential realized in single
heterojunctions. Further studies are needed to provide a basis for
accurate comparisons with experimental data, as the actual shape of
the confinement potential and the (here neglected) anisotropy of the
dispersion may influence the exact functional dependence of the
exchange energy on characteristic system parameters such as the
2D-hole sheet density.

\begin{acknowledgments}
  Work at Argonne was supported by DOE BES under Contract No.\
  DE-AC02-06CH11357.
\end{acknowledgments}

%

\end{document}